\documentclass[11pt]{article}
\usepackage{hyperref}
\pdfoutput=1
\begin{document}
\title{Particle Jet Formation During Explosive Dispersal of Solid Particles }
\author{David L. Frost$^{1}$, Yann Gr$\acute{e}$goire$^{2}$, Oren Petel$^{1}$, Samuel Goroshin$^{1}$, Fan Zhang$^{3}$ \\
\\\vspace{6pt} $^{1}$Mechanical Engineering Department, \\ McGill University, Montreal, QC H3A 2K6, Canada \\
\\\vspace{6pt} $^{2}$Institut Pprime, CNRS, ENSMA, \\ Futuroscope-Chasseneuil, France \\
\\\vspace{6pt} $^{3}$Defense Research and Development Canada - Suffield, \\ Medicine Hat, AB T1A 8K6, Canada}
\maketitle
\begin{abstract}
Previous experimental studies have shown that when a layer of solid particles is explosively dispersed, the particles often develop a non-uniform spatial distribution.  The instabilities within the particle bed and at the particle layer interface likely form on the timescale of the shock propagation through the particles.  The mesoscale perturbations are manifested at later times in experiments by the formation of coherent clusters of particles or jet-like particle structures, which are aerodynamically stable.  A number of different mechanisms likely contribute to the jet formation including shock fracturing of the particle bed and particle-particle interactions in the early stages of the dense gas-particle flow.  Aerodynamic wake effects at later times contribute to maintaining the stability of the jets. The experiments shown in this fluid dynamics video were carried out in either spherical or cylindrical geometry and illustrate the formation of particle jets during the explosive dispersal process.    The number of jet-like structures that are generated during the dispersal of a dry powder bed is compared with the number formed during the dispersal of the same volume of water.  The liquid dispersal generates a larger number of jets, but they fragment and dissipate sooner.  When the particle bed is saturated with water and explosively dispersed, the number of particle jets formed is larger than both the dry powder and pure water charges.   This effect is particularly evident in cylindrical geometry where the number of jets formed for a saturated particle bed is about an order of magnitude larger than for a dry particle bed.  Further experiments on jet formation with other particles in a conical geometry have been reported in an earlier publication$^{1}$.  In this paper, it was found that the number of particle jets that form tends to scale with a particle compaction Reynolds number corresponding to the ratio of inertial to frictional forces of the particle system.
\end{abstract}
\section{Experimental Details}
The experiments shown in the video were carried out in two different geometries.  The casings for the spherical charges were prepared by removing the filament from thin-walled globe light bulbs with a diameter of about 12.5 cm and a volume about 1 liter.  The particles were dispersed with a centrally located spherical booster charge of 28 g of C4 explosive contained within a plastic sphere and initiated with a detonator.  The charge with the dry glass beads contained 1,332 g of 120 $\mu$m spherical glass beads (size 10 ballotini impact spheres from Potters Industries).  The charge with pure water contained 902 g of water whereas the charge with the wet particle bed contained 1,482 g of the same glass spheres saturated with 348 g of water.  The cylindrical charges consisted of a 10-cm dia glass cylinder (30 cm long) filled with particles and with a central detonating cord (80 g PETN/m) running the length of the cylinder.  A 2-cm wide section of size 10 glass beads was surrounded by two 14-cm wide layers of 200 $\mu$m steel beads for confinement.  All of the experiments were visualized with a Photron SA5 high-speed video camera operating typically at 10,000 fr/s.  The cylindrical tests were viewed from the end with the line of sight to the camera parallel to the cylinder axis.  A layer of shaving foam was placed on the end of the charge to eliminate the flash from the detonating cord.  Expansion of the shaving cream is also visible on the high-speed videos. 
\section{Acknowledgements} 
Support with the experiments from Rick Guilbeault of the Canadian Explosive Research Centre is gratefully acknowledged.  Assistance with the experiments was also provided by Alexandre Coderre-Chabot, Melanie Tetrault-Friend, Francois-Xavier Jett$\acute{e}$, and Oren Petel.  Support for the work was provided by Defense Research and Development Canada - Suffield and the U.S. Defence Threat Reduction Agency. Yann Gr$\acute{e}$goire was supported under a postdoctoral fellowship provided by Division G$\acute{e}$n$\acute{e}$rale de L'Armement of France.   
\\
\\
$^{1}$Frost, D.L., Gr$\acute{e}$goire, Y, Goroshin, S., and Zhang, F., ''Interfacial instabilities in explosive gas-particle flows,'' Proceedings of 23$^{rd}$ International Colloquium on the Dynamics of Explosions and Reactive Systems, Irvine, CA, July 24--29, 2011.
\end{document}